\documentclass[
 aip,
 amsmath,amssymb,
 preprint,
]{revtex4-1}

\usepackage{graphicx}
\usepackage{dcolumn}
\usepackage{comment}
\usepackage{bm}
\usepackage[utf8]{inputenc}
\usepackage[T1]{fontenc}
\usepackage{mathptmx}
\usepackage{etoolbox}
\usepackage{upgreek}
\usepackage{xcolor}
\usepackage{hyperref}
\usepackage{stmaryrd}
\usepackage{subfigure}

\makeatletter
\def\@email#1#2{%
 \endgroup
 \patchcmd{\titleblock@produce}
  {\frontmatter@RRAPformat}
  {\frontmatter@RRAPformat{\produce@RRAP{*#1\href{mailto:#2}{#2}}}\frontmatter@RRAPformat}
  {}{}
}%
\makeatother

\begin{document}

\title{Rankine-Hugoniot conditions in Q-variables: a wave-aligned formulation of MHD discontinuities}

\author{Anna Krupka}
\affiliation{Centre for mathematical Plasma Astrophysics, Department of Mathematics, KU Leuven, Celestijnenlaan 200B, B-3001 Leuven, Belgium}
\email{anna.krupka@kuleuven.be}

\author{Tijs Van Hoof}
\affiliation{Department of Mathematics, KU Leuven Campus Kortrijk, Etienne Sabbelaan 53, bus 2400, 8500 Kortrijk, Belgium}

\author{Tom Van Doorsselaere}
\affiliation{Centre for mathematical Plasma Astrophysics, Department of Mathematics, KU Leuven, Celestijnenlaan 200B, B-3001 Leuven, Belgium}

\begin{abstract}
The recently developed Q-variable formalism generalises the Elsässer representation by providing a wave-aligned representation applicable to a broad class of magnetohydrodynamic disturbances, including Alfvénic, fast, slow, and kink waves. While this framework has proven useful for the study of wave dynamics and turbulence, its behaviour in the presence of plasma discontinuities has not yet been established. In this work, we derive the complete set of Rankine-Hugoniot jump conditions in terms of the Q-variables by rewriting the ideal MHD equations in a form suitable for shock-frame jump analysis. This yields explicit jump relations for mass, momentum, magnetic flux, and energy. We then demonstrate analytically that these relations are exactly equivalent to the classical MHD Rankine-Hugoniot conditions. This reformulation provides a wave-aligned representation of MHD discontinuities and offers a natural framework for discussing directional wave content and branch-restricted limits when $\alpha$, the wave-branch parameter entering the Q-variable definition, is chosen consistently with the relevant characteristic speed. The resulting formulation is well suited for the analysis of wave-shock interactions in magnetised plasmas, with potential applications to the solar wind, magnetospheric systems, and large-scale models of structured plasma environments such as UAWSOM.

\end{abstract}
\maketitle

\section{Introduction}

Discontinuities such as shocks, contact surfaces, and rotational or tangential discontinuities are fundamental structures in space and astrophysical plasmas. They arise naturally in a wide range of environments, including the solar wind \citep{Bruno_Carbone_2013},
coronal mass ejection (CME)-driven shocks \citep{Vrsnak_2008},
the Earth’s bow shock and magnetosheath \citep{Treumann_2009},
and the heliospheric termination and bow shocks \citep{Stone_2005, Stone_2008, Baranov_1971}.
In the solar corona, current sheets and tangential discontinuities are frequently identified, particularly within magnetic reconnection regions and along streamer boundaries \citep{Priest_Forbes_2000}. These structures govern the conversion of bulk-flow energy to heat, modulate the transport of momentum and magnetic flux, and shape the global dynamics of heliospheric and broader astrophysical environments. From a wave-dynamical perspective, shocks arise from the nonlinear steepening of characteristic MHD modes. A formulation based directly on wave variables therefore provides a physically transparent framework for analysing their structure and dynamics.

Across such discontinuities, the physical relations between upstream and downstream states are governed by the classical Rankine-Hugoniot (RH) conditions. These jump conditions may be derived either by integrating the local conservation laws of mass, momentum, magnetic flux, and energy across an infinitesimally thin interface, or equivalently by applying the corresponding conservation laws in integral form to a control volume enclosing the discontinuity \citep{deHoffmann_Teller_1950, Jeffrey_Taniuti_2000, Landau_Lifshitz_1984}. The RH conditions form the basis of magnetohydrodynamic (MHD) shock theory and govern our understanding of wave-shock interactions, heating, and particle acceleration.

Despite their fundamental role, the RH conditions are traditionally expressed in terms of the primitive MHD variables $(\rho, \boldsymbol{V}, p, \boldsymbol{B})$. While complete, this formulation does not naturally separate the contributions of different MHD wave modes. This becomes limiting when examining shocks interacting with turbulence or when analysing how individual modes, such as Alfvén, fast, or slow waves, propagate through or interact with shocks.

A major step forward was the introduction of the Elsässer variables
\citep{Elsasser_1950}
\[
\boldsymbol{Z}^{\pm} = \boldsymbol{V} \pm \boldsymbol{V}_{A},
\]
where $\boldsymbol{V}$ is the plasma velocity and
\[
\boldsymbol{V}_{A}
=
\frac{\boldsymbol{B}}{\sqrt{\mu\rho}}
\]
is the Alfvén velocity, with $\boldsymbol{B}$ the magnetic field, $\rho$ the plasma mass density, and $\mu$ the magnetic permeability.

These variables isolate the Alfvénic dynamics of the MHD equations and form the basis for modern theories of Alfvénic turbulence in the solar wind \citep{Bruno_Carbone_2013}. Because of their natural separation of oppositely propagating Alfvén waves, Elsässer variables have been widely used to describe solar-wind turbulence and its nonlinear cascade \citep{Dobrowolny_1980, Marsch_Tu_1989, Tu_Marsch_Thieme_1989, Velli_Grappin_Mangeney_1989, Zhou_Matthaeus_1989, Grappin_1990}. Their success motivated attempts to reformulate the full MHD system in terms of Elsässer variables and density alone \citep{Marsch_Mangeney_1987}.

However, Elsässer variables are intrinsically tied to the Alfvén speed and therefore do not cleanly represent other wave modes. Slow and fast magnetosonic waves necessarily involve mixed contributions from $\boldsymbol{Z}^{+}$ and $\boldsymbol{Z}^{-}$ even in a homogeneous plasma \citep{Magyar_2019}, and this behaviour persists in non-uniform plasmas \citep{Ismayilli_2022}. More generally, attempts to extend the Elsässer formalism beyond purely Alfvénic dynamics encounter intrinsic theoretical limitations \citep{Galtier_2023}, which restrict the ability of Elsässer-based descriptions to capture the full wave content and mode coupling present in magnetised plasmas \citep{Schekochihin_2009}. These limitations motivate the development of wave variables that treat all MHD modes in a unified manner, provide a representation aligned with propagation directions, and remain valid in inhomogeneous plasmas.

To address this, \citet{VanDoorsselaere_2024} introduced the Q-variable formalism, a generalisation of Elsässer variables defined as
\[
    \boldsymbol{Q}^{\pm} = \boldsymbol{V} \pm \alpha \boldsymbol{B},
\]
where $\alpha$ is a parameter related to the phase speed of a propagating MHD disturbance. The parameter $\alpha$ is not an independent variable but is determined by the local plasma state and the chosen wave mode, and therefore varies consistently with the underlying MHD variables. Recent work has shown that Q-variables provide a unified framework for describing Alfvén, kink, and magnetoacoustic waves and allow their dispersion relations to be derived analytically \citep{VanDoorsselaere_2024}.

The goal of the present work is to extend the Q-variable formalism to the regime of plasma discontinuities by deriving the full set of Rankine-Hugoniot conditions in terms of $\boldsymbol{Q}^{\pm}$. This allows us to (i) test whether Q-variables form a complete and self-consistent representation of MHD; (ii) demonstrate that the Q-based RH system exactly reproduces the classical MHD jump conditions; and (iii) interpret the jump conditions in terms of directional characteristic contributions. When $\alpha$ is chosen consistently with the relevant characteristic speed on each side of the discontinuity, the variables $\boldsymbol{Q}^{\pm}$ provide a representation aligned with oppositely directed propagation branches. This provides a framework for analysing how directional wave components contribute to the structure and dynamics of MHD discontinuities, which are not explicitly separated in the classical formulation.

In this work, we demonstrate that the Q-variable formalism is algebraically equivalent to classical MHD at the level of the full Rankine–Hugoniot system, while simultaneously providing a wave-aligned reorganisation of the dynamics. Beyond this equivalence, the Q-variable framework offers a structural advantage: it expresses the jump conditions directly in terms of quantities aligned with characteristic propagation directions. This makes it possible to express the velocity--magnetic-field part of the jump conditions in terms of contributions associated with oppositely directed characteristic branches, without requiring an additional decomposition at the level of the classical variables. In the classical formulation, such directional information is not explicitly separated and must instead be reconstructed indirectly. This representation therefore provides a natural basis for analysing wave-discontinuity interactions and directional coupling at shocks, with potential applications in turbulence modelling and large-scale solar-wind simulations, including models such as UAWSoM~\cite{McMurdo_2026}.

This paper is organised as follows.
Section~\ref{sec:Qframework} introduces the Q-variable formalism and the discontinuity setup.
Section~\ref{sec:QRH} derives the Rankine-Hugoniot conditions in Q-variables and demonstrates their equivalence with classical MHD.
Section~\ref{sec:interpretation} discusses the physical interpretation of the results, including directional propagation and limiting cases.
Finally, Section~\ref{sec:conclusions} summarises the main findings and outlines future applications.

\section{Q-variable formalism and shock framework}
\label{sec:Qframework}
\subsection{Q-variable formalism of MHD}

The Q-variable formalism employs two wave-frame variables
\begin{equation}
    \boldsymbol{Q}^{\pm} = \boldsymbol{V} \pm \alpha\, \boldsymbol{B},
\end{equation}
where $\alpha$ is a parameter associated with the phase speed of a general wave, and 
$\boldsymbol{V}$ and $\boldsymbol{B}$ denote the plasma velocity and magnetic field, respectively.
The inverse relations are
\begin{equation}
\label{V_and_B}
    \boldsymbol{V} = \frac{\boldsymbol{Q}^+ + \boldsymbol{Q}^-}{2},
    \qquad
    \boldsymbol{B} = \frac{\boldsymbol{Q}^+ - \boldsymbol{Q}^-}{2\alpha}.
\end{equation}
The derivatives comoving with the $\pm$ propagation directions are defined as
\begin{equation}
    \frac{D^\pm}{Dt}
    = \frac{\partial}{\partial t} + \boldsymbol{Q}^{\pm}\cdot\nabla .
\end{equation}

When $\alpha = 1/\sqrt{\mu\rho}$, the Q-variables reduce to the classical Elsässer variables. In the general case, the choice of $\alpha$ determines which wave mode is separated into upward and downward propagating components. The parameter $\alpha$ is not an independent dynamical variable but is determined by the local plasma state and the chosen wave mode. It may therefore vary across a discontinuity through its dependence on the underlying MHD variables, but it does not satisfy a separate Rankine-Hugoniot condition. Instead, it is evaluated locally from the plasma state and the selected wave branch, so that its upstream and downstream values are related implicitly through the jump conditions for the underlying MHD variables.

The ideal MHD equations can be rewritten entirely in terms of $\boldsymbol{Q}^{\pm}$, as derived by \cite{VanDoorsselaere_2024}. The system consists of a modified solenoidal constraint, two mass-conservation equations, and two momentum equations for $\boldsymbol{Q}^{\pm}$.  
The solenoidal constraint becomes
\begin{equation}
\label{eq:solenoidal}
    \nabla\cdot(\boldsymbol{Q}^+ - \boldsymbol{Q}^-)
    - (\boldsymbol{Q}^+ - \boldsymbol{Q}^-)\cdot\nabla\ln\alpha = 0.
\end{equation}
Mass conservation in each characteristic frame reads
\begin{equation}
\label{eq:mass}
    \frac{D^\pm}{Dt}\,\ln\rho
    = -\frac{1}{2}\nabla\cdot(3\boldsymbol{Q}^{\pm} - \boldsymbol{Q}^{\mp})
      \pm \frac{\boldsymbol{Q}^+ - \boldsymbol{Q}^-}{2}\cdot\nabla\ln(\rho\alpha^{2}).
\end{equation}
The momentum equations for $\boldsymbol{Q}^{\pm}$ are
\begin{align}
\label{eq:momentum}
    &\frac{D^{\mp}}{Dt}\boldsymbol{Q}^{\pm}
    \mp \frac{\boldsymbol{Q}^+ - \boldsymbol{Q}^-}{4}
        \frac{D^{\mp}}{Dt}\ln(\rho\alpha^2)
    = -v_s^2 \nabla\ln\rho
      - \frac{1}{8}\!\left(1 - \frac{\Delta\alpha^2}{\alpha^2}\right)
        \nabla(\boldsymbol{Q}^+ - \boldsymbol{Q}^-)^2 
        \nonumber\\[0.3em]
    &\quad + \frac{1}{4}\!\left(1 - \frac{\Delta\alpha^2}{\alpha^2}\right)
        (\boldsymbol{Q}^+ - \boldsymbol{Q}^-)^2 \nabla\ln\alpha 
      - \frac{1}{4}\frac{\Delta\alpha^2}{\alpha^2}
        (\boldsymbol{Q}^+ - \boldsymbol{Q}^-)\cdot\nabla(\boldsymbol{Q}^+ - \boldsymbol{Q}^-)
        \nonumber\\[0.3em]
    &\quad + \frac{1}{4}\frac{\Delta\alpha^2}{\alpha^2}
        (\boldsymbol{Q}^+ - \boldsymbol{Q}^-)\nabla\cdot(\boldsymbol{Q}^+ - \boldsymbol{Q}^-)
      \pm \frac{\boldsymbol{Q}^+ - \boldsymbol{Q}^-}{4}
        \frac{D^{\pm}}{Dt}\ln(\rho\alpha^2).
\end{align}

Here $v_s = \sqrt{\gamma p/\rho}$ denotes the adiabatic sound speed of the plasma, and $\Delta\alpha^2 \equiv \alpha^2-1/(\mu\rho)$ denotes the deviation of the wave parameter from its Alfv\'en-wave value. This quantity may be either positive or negative, depending on whether the corresponding phase speed is greater or smaller than the Alfv\'en speed, and should not be confused with a differential operator acting on $\alpha^2$. By assuming an adiabatic equation of state, $p = p_0 (\rho / \rho_0)^\gamma$, the system of equations is closed. The Q-variable energy equation is introduced later in the Rankine-Hugoniot analysis, where it is obtained by rewriting the classical MHD energy conservation law in terms of the Q-variables.

The Q-MHD system above provides a complete formulation of the ideal MHD equations that is equivalent to the classical description. In the next subsection, we introduce the shock-frame geometry and the limiting identities required to evaluate the behaviour of each Q-MHD term across an infinitesimally thin discontinuity.

\subsection{Shock geometry and derivative limits}
\label{shockgeometry}

To derive the Rankine-Hugoniot conditions for the Q-variable system, we first review the standard geometric setup for an infinitesimally thin shock and the behaviour of derivatives across such a discontinuity.

\begin{figure}[ht]
    \centering
    \includegraphics[width=0.55\textwidth]{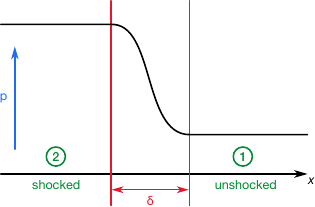}
    \caption{Schematic shock geometry and notation. The schematic pressure increase indicates a sudden jump in pressure associated with the shock. The unit normal $\mathbf{n}$ points from the downstream (2) to the upstream (1) state. A quantity $f$ has a jump $\llbracket f \rrbracket = f_1-f_2$. In the laboratory frame, the discontinuity propagates along $\mathbf{n}$ with shock speed $u$. The parameter $\delta$ denotes the schematic thickness of the transition layer, which is taken to zero in the Rankine-Hugoniot limit.}
    \label{fig:shockgeometry}
\end{figure}

We consider a steady, planar shock separating an upstream (unshocked) region~1 from a downstream (shocked) region~2, using standard MHD shock notation and sign conventions \cite{Goedbloed_Keppens_Poedts_2019}. The unit normal vector $\mathbf{n}$ is taken to point from region~2 toward region~1. Figure~\ref{fig:shockgeometry} summarises the setup and notation.

The jump of any quantity $f$ is defined as
\begin{equation}
    \llbracket f \rrbracket \equiv f_{1} - f_{2},
\end{equation}
where the subscripts refer to the upstream and downstream values.

In the limit where the shock thickness $\delta \to 0$, variations normal to the surface dominate over tangential gradients. Introducing a coordinate $l$ measured along the normal and increasing from region~2 to region~1, we have $\partial f / \partial l \to \infty$ within the thin transition layer, while $f$ remains finite on either side. Integrating spatial derivatives across this layer yields the standard identities used in shock derivations.

For any sufficiently smooth scalar $f$, integrating $\nabla f$ across the discontinuity gives
\begin{equation}
\label{eq:nabla_limit}
    \lim_{\delta \to 0} \int_{1}^{2} \nabla f\, dl
    = - \mathbf{n} \lim_{\delta \to 0} \int_{1}^{2} \frac{\partial f}{\partial l}\, dl
    = \mathbf{n}\,\llbracket f\rrbracket.
\end{equation}
Because the integration proceeds from region~1 to region~2, opposite to the orientation of $\mathbf{n}$, the sign appears as written.

In the laboratory frame, time derivatives arise from the motion of the discontinuity relative to the plasma. If the shock moves with normal speed $u$ (positive toward region~1), then
\begin{equation}
\label{eq:time_limit}
    \lim_{\delta \to 0} \int_{1}^{2} \frac{\partial f}{\partial t}\, dl
    = \lim_{\delta \to 0} \int_{1}^{2} \frac{\partial f}{\partial l}\frac{\partial l}{\partial t}\, dl
    = -u\,\llbracket f\rrbracket.
\end{equation}

Similarly, flux-divergence terms produce
\begin{equation}
\label{eq:div_limit}
    \lim_{\delta \to 0} \int_{1}^{2} \nabla \cdot \boldsymbol{F} \, dl
        = \mathbf{n} \cdot \llbracket \boldsymbol{F} \rrbracket.
\end{equation}

These relations serve as the starting point for determining the Rankine-Hugoniot conditions in any variable set. For the Q-variable system, the derivative of primary interest is the wave-frame derivative
\begin{equation}
    \frac{D^\pm f}{Dt}
        = \frac{\partial f}{\partial t}
          + \boldsymbol{Q}^{\pm} \cdot \nabla f.
\end{equation}

To determine its behaviour across a shock, we apply the Reynolds transport theorem \citep{Leal_2007}, which gives
\begin{equation}
    \int_{1}^{2} \boldsymbol{Q}^{\pm} \cdot \nabla f \, dl
        = - \left(\boldsymbol{Q}^{\pm} \cdot \mathbf{n}\right) \int_{1}^{2} \frac{\partial f}{\partial l}\, dl .
\end{equation}
Using \eqref{eq:nabla_limit}--\eqref{eq:time_limit}, we obtain
\begin{equation}
\label{eq:limit_Dpm}
    \lim_{\delta \to 0} \int_{1}^{2} \frac{D^\pm f}{D t}\, dl
    = -u\,\llbracket f\rrbracket
      + (\boldsymbol{Q}^{\pm}\cdot\mathbf{n})\,\llbracket f\rrbracket.
\end{equation}

For later convenience, we introduce the normal and tangential components of the Q-variables,
\begin{equation}
    Q^{\pm}_{n} = \boldsymbol{Q}^{\pm} \cdot \mathbf{n},
    \qquad
    \boldsymbol{Q}^{\pm}_{t}
        = \boldsymbol{Q}^{\pm} - Q^{\pm}_{n} \mathbf{n}.
\end{equation}
It is also convenient to work in the shock frame, in which the discontinuity is stationary. In this frame, the transformed Q-variables are
\begin{equation}
    \boldsymbol{Q}^{\prime\pm} = \boldsymbol{Q}^{\pm} - u \mathbf{n},
\end{equation}
so that
\begin{equation}
    Q^{\prime\pm}_{n} = Q^{\pm}_{n} - u,
    \qquad
    \boldsymbol{Q}^{\prime\pm}_{t} = \boldsymbol{Q}^{\pm}_{t}.
\end{equation}
Expression~\eqref{eq:limit_Dpm} therefore reduces to
\begin{equation}
\label{eq:shock_Dpm}
    \frac{D^\pm f}{Dt} \;\longrightarrow\; 
        Q^{\prime\pm}_{n}\, \llbracket f \rrbracket,
\end{equation}
which is the substitution rule used when deriving the Q-variable Rankine-Hugoniot relations.

Collecting the results above, the shock-limit replacements for the differential operators are
\begin{align}
    \frac{\partial f}{\partial t}
        &\;\longrightarrow\; -u\, \llbracket f \rrbracket, \\
    \nabla f
        &\;\longrightarrow\; \mathbf{n}\, \llbracket f \rrbracket, \\
    \nabla \cdot \boldsymbol{F}
        &\;\longrightarrow\; \mathbf{n} \cdot \llbracket \boldsymbol{F} \rrbracket, \\
    \boldsymbol{Q}^{\pm} \cdot \nabla f
        &\;\longrightarrow\; Q^{\pm}_{n}\, \llbracket f \rrbracket, \\
    \frac{D^\pm f}{Dt}
        &\;\longrightarrow\; Q^{\prime\pm}_{n}\, \llbracket f \rrbracket.
\end{align}

These identities form the basis for computing the Rankine-Hugoniot conditions in terms of the Q-variables, which we derive in the following section.

\section{Rankine-Hugoniot conditions in Q-variables}
\label{sec:QRH}
\subsection{Derivation of Rankine-Hugoniot conditions in the Q-formalism}

Using the Q-MHD system \eqref{eq:solenoidal}--\eqref{eq:momentum} and the shock-limit identities introduced in Sec.~\ref{sec:Qframework}, we derive the Rankine-Hugoniot conditions in terms of the Q-variables. Although the derivation proceeds through the Q-variable system, the resulting jump relations are verified explicitly below to reproduce the classical conservative MHD Rankine-Hugoniot conditions exactly.
We proceed equation by equation, rewriting each relation in a form suitable for applying the substitutions
\[
    \nabla f \;\rightarrow\; \mathbf{n}\,\llbracket f \rrbracket,
    \qquad
    \nabla \cdot \boldsymbol{F} \;\rightarrow\; \mathbf{n}\cdot \llbracket \boldsymbol{F} \rrbracket,
\]
and, where appropriate,
\[
    \frac{D^\pm f}{Dt} \;\rightarrow\; Q^{\prime\pm}_n \llbracket f \rrbracket,
\]
where $Q^{\prime\pm}_n$ denotes the normal components of the Q-variables in the shock frame.

\subsubsection{Normal magnetic flux}
\label{sec:RH_normal_flux}

We begin with the solenoidal constraint of the Q-system,
\begin{equation}
\label{eq:RH_solenoidal_start}
    \nabla\cdot(\boldsymbol{Q}^+ - \boldsymbol{Q}^-)
    - (\boldsymbol{Q}^+ - \boldsymbol{Q}^-)\cdot\nabla\ln\alpha = 0.
\end{equation}
It is convenient to rewrite this equation so that $\alpha$ appears explicitly. Using $\nabla\ln\alpha = (1/\alpha)\,\nabla\alpha$, we obtain
\begin{equation}
    \nabla\cdot(\boldsymbol{Q}^+ - \boldsymbol{Q}^-)
    - \frac{1}{\alpha} (\boldsymbol{Q}^+ - \boldsymbol{Q}^-)\cdot\nabla\alpha = 0.
\end{equation}
Multiplying by $1/\alpha$ and applying the product rule yields
\begin{equation}
    \nabla\cdot\!\left(
        \frac{\boldsymbol{Q}^+ - \boldsymbol{Q}^-}{\alpha}
    \right) = 0.
\end{equation}

This form is directly suitable for applying the shock-limit substitution for a divergence. Integrating across the shock layer and using \eqref{eq:div_limit}, we obtain
\begin{equation}
    \mathbf{n}\cdot
    \bigg\llbracket
        \frac{\boldsymbol{Q}^+ - \boldsymbol{Q}^-}{\alpha}
    \bigg\rrbracket = 0,
\end{equation}
or, in terms of the normal components,
\begin{equation}
\label{eq:RH_normal_flux_Q}
    \bigg\llbracket
        \frac{Q^+_{n} - Q^-_{n}}{\alpha}
    \bigg\rrbracket = 0.
\end{equation}

Transforming to the shock frame does not change this expression, since $Q^{\pm}_{n} = Q^{\prime\pm}_{n} + u$, and their difference is invariant, $Q^+_{n} - Q^-_{n} = Q^{\prime+}_{n} - Q^{\prime-}_{n}$. Therefore, the RH condition can be written equivalently as
\begin{equation}
\label{eq:RH1}
    \bigg\llbracket
        \frac{Q^{\prime+}_{n} - Q^{\prime-}_{n}}{\alpha}
    \bigg\rrbracket = 0.
\end{equation}

Equation~\eqref{eq:RH1} is the first Q-variable Rankine-Hugoniot condition and expresses the continuity of the normal magnetic flux in the Q-formalism.

\subsubsection{Mass conservation}
\label{sec:RH_mass}

The next jump condition follows from the Q-form of mass conservation, equation~\eqref{eq:mass}.  
Adding the $+$ and $-$ equations eliminates the explicit coupling terms proportional to $\boldsymbol{Q}^+ - \boldsymbol{Q}^-$ and yields
\begin{equation}
\label{eq:mass_plusminus}
    \frac{D^+}{Dt}\ln\rho
    + \frac{D^-}{Dt}\ln\rho
    = - \nabla\cdot(\boldsymbol{Q}^+ + \boldsymbol{Q}^-).
\end{equation}
To obtain a conservative form, we multiply by $\rho$ and expand the comoving derivatives, so that the left-hand side becomes
\begin{align}
     \rho\left(
        \frac{1}{\rho}\frac{D^+ \rho}{Dt}
      + \frac{1}{\rho}\frac{D^- \rho}{Dt}
    \right)
    = 2\frac{\partial\rho}{\partial t}
       + (\boldsymbol{Q}^+ + \boldsymbol{Q}^-)\cdot\nabla\rho.
\end{align}
Equation~\eqref{eq:mass_plusminus} thus becomes
\begin{equation}
    2\frac{\partial\rho}{\partial t}
    + (\boldsymbol{Q}^+ + \boldsymbol{Q}^-)\cdot\nabla\rho
    + \rho\,\nabla\cdot(\boldsymbol{Q}^+ + \boldsymbol{Q}^-)
    = 0.
\end{equation}
Using the product rule, this can be written in conservative form as
\begin{equation}
\label{eq:rho_conservative}
    2\frac{\partial\rho}{\partial t}
    + \nabla\cdot\bigl[\rho(\boldsymbol{Q}^+ + \boldsymbol{Q}^-)\bigr]
    = 0.
\end{equation}

Applying the shock-limit substitutions for the time derivative and the divergence, and using $Q_n^{\pm} = \boldsymbol{Q}^{\pm}\cdot\mathbf{n}$, we obtain
\begin{equation}
    -2u\,\llbracket \rho \rrbracket
    + \biggl\llbracket
        \rho\,(Q_n^{+} + Q_n^{-})
      \biggr\rrbracket
    = 0.
\end{equation}

We now express this condition in the shock frame.  
With $Q_n^{\prime\pm} = Q_n^{\pm} - u$, we have $Q_n^{+} + Q_n^{-}
    = Q_n^{\prime+} + Q_n^{\prime-} + 2u$, and thus
\begin{align}
    \biggl\llbracket
        \rho\,(Q_n^{+} + Q_n^{-})
    \biggr\rrbracket
    &= \biggl\llbracket
        \rho\,(Q_n^{\prime+} + Q_n^{\prime-})
      \biggr\rrbracket
       + 2u\,\llbracket \rho \rrbracket.
\end{align}
Substituting this into the previous relation cancels the terms proportional to $\llbracket \rho \rrbracket$ and yields
\begin{equation}
\label{eq:RH2}
    \biggl\llbracket
        \rho\,(Q_n^{\prime+} + Q_n^{\prime-})
    \biggr\rrbracket
    = 0.
\end{equation}

Equation~\eqref{eq:RH2} is the Q-variable form of mass conservation across the shock and expresses the continuity of the normal mass flux in terms of the shock-frame Q-variables.

\subsubsection{Tangential magnetic flux}

We now derive the third Rankine-Hugoniot condition by subtracting the two momentum equations for $\boldsymbol{Q}^{+}$ and $\boldsymbol{Q}^{-}$. Subtracting equation~\eqref{eq:momentum} written for the $+$ and $-$ branches yields
\begin{equation}
\label{eq:raw_diff}
    \frac{D^-}{Dt}\boldsymbol{Q}^{+}
    - \frac{D^+}{Dt}\boldsymbol{Q}^{-}
    - \left(\frac{\boldsymbol{Q}^{+}-\boldsymbol{Q}^{-}}{4}\right)
        \left(
            \frac{D^-}{Dt}\ln(\rho\alpha^{2})
            +
            \frac{D^+}{Dt}\ln(\rho\alpha^{2})
            + \frac{D^-}{Dt}\ln\rho
            + \frac{D^+}{Dt}\ln\rho
        \right)
    = 0.
\end{equation}

Using the logarithmic identities
$\ln(xy)=\ln x + \ln y$ and $\ln(x^{a})=a\ln x$, this becomes
\begin{equation}
\label{eq:expanded_logs}
    \frac{D^-}{Dt}\boldsymbol{Q}^{+}
    - \frac{D^+}{Dt}\boldsymbol{Q}^{-}
    - \left(\frac{\boldsymbol{Q}^{+}-\boldsymbol{Q}^{-}}{2}\right)
        \left(
            \frac{D^-}{Dt}\ln\alpha
            + \frac{D^+}{Dt}\ln\alpha
            + \frac{D^-}{Dt}\ln\rho
            + \frac{D^+}{Dt}\ln\rho
        \right)
    = 0.
\end{equation}

We now use the sum of the two mass equations, obtained previously in
equation~\eqref{eq:mass_plusminus}, and multiply the entire equation by $2$ to obtain
\begin{align}
\label{eq:after_mass}
    &2\frac{D^-}{Dt}\boldsymbol{Q}^{+}
    - 2\frac{D^+}{Dt}\boldsymbol{Q}^{-}
    + \left(\boldsymbol{Q}^{+}-\boldsymbol{Q}^{-}\right)
      \nabla\!\cdot(\boldsymbol{Q}^{+}+\boldsymbol{Q}^{-}) \nonumber \\
    &\quad - \left(\boldsymbol{Q}^{+}-\boldsymbol{Q}^{-}\right)
      \left(\frac{D^-}{Dt}\ln\alpha + \frac{D^+}{Dt}\ln\alpha\right)
    = 0.
\end{align}

Expanding the comoving derivatives and regrouping terms gives
\begin{align}
\label{eq:expanded_predivide}
    &2\frac{\partial}{\partial t}(\boldsymbol{Q}^{+}-\boldsymbol{Q}^{-})
      - (\boldsymbol{Q}^{+}-\boldsymbol{Q}^{-})\,\frac{2}{\alpha}
        \frac{\partial\alpha}{\partial t}
      + 2\boldsymbol{Q}^{-}\!\cdot\nabla\boldsymbol{Q}^{+}
      - 2\boldsymbol{Q}^{+}\!\cdot\nabla\boldsymbol{Q}^{-} \nonumber\\
    &\quad
      +(\boldsymbol{Q}^{+}-\boldsymbol{Q}^{-})
         \nabla\!\cdot(\boldsymbol{Q}^{+}+\boldsymbol{Q}^{-})
      -(\boldsymbol{Q}^{+}-\boldsymbol{Q}^{-})
         \frac{1}{\alpha}(\boldsymbol{Q}^{+}+\boldsymbol{Q}^{-})
         \!\cdot\nabla\alpha
      = 0 .
\end{align}

Multiplying equation~\eqref{eq:expanded_predivide} by $1/\alpha$ and using the quotient rule on the first two terms and on the last two terms yields
\begin{align}
\label{eq:after_divide}
    &2\frac{\partial}{\partial t}
        \left(\frac{\boldsymbol{Q}^{+}-\boldsymbol{Q}^{-}}{\alpha}\right)
      + \frac{2}{\alpha}
        \left(\boldsymbol{Q}^{-}\!\cdot\nabla\boldsymbol{Q}^{+}
             - \boldsymbol{Q}^{+}\!\cdot\nabla\boldsymbol{Q}^{-}\right)
      \nonumber\\
    &\qquad
      + (\boldsymbol{Q}^{+}-\boldsymbol{Q}^{-})
        \nabla\!\cdot\left(
            \frac{\boldsymbol{Q}^{+}+\boldsymbol{Q}^{-}}{\alpha}
        \right)
      = 0.
\end{align}

We now rewrite the last term using the identity for the divergence of a
dyadic product,
\[
\mathbf{B}\,\nabla\!\cdot\mathbf{A}
  = \nabla\!\cdot(\mathbf{A}\mathbf{B})
    - (\mathbf{A}\!\cdot\nabla)\mathbf{B},
\]
which gives
\begin{align}
\label{eq:prejump_divergence_form}
    &2\frac{\partial}{\partial t}
        \left(\frac{\boldsymbol{Q}^{+}-\boldsymbol{Q}^{-}}{\alpha}\right)
      + \frac{2}{\alpha}
        \left(\boldsymbol{Q}^{-}\!\cdot\nabla\boldsymbol{Q}^{+}
             - \boldsymbol{Q}^{+}\!\cdot\nabla\boldsymbol{Q}^{-}\right)
      \nonumber\\
    &\qquad
      + \nabla\!\cdot\!
        \left[
            \left(\frac{\boldsymbol{Q}^{+}+\boldsymbol{Q}^{-}}{\alpha}\right)
            (\boldsymbol{Q}^{+}-\boldsymbol{Q}^{-})
        \right]
      - \left(\frac{\boldsymbol{Q}^{+}+\boldsymbol{Q}^{-}}{\alpha}\right)
        \!\cdot\nabla(\boldsymbol{Q}^{+}-\boldsymbol{Q}^{-})
      =0 .
\end{align}

Combining the second and fourth terms yields
\begin{align}
\label{eq:prejump_final}
    &2\frac{\partial}{\partial t}
        \left(\frac{\boldsymbol{Q}^{+}-\boldsymbol{Q}^{-}}{\alpha}\right)
      + \nabla\!\cdot\!
        \left[
            \left(\frac{\boldsymbol{Q}^{+}+\boldsymbol{Q}^{-}}{\alpha}\right)
            (\boldsymbol{Q}^{+}-\boldsymbol{Q}^{-})
        \right]
      - \left(\frac{\boldsymbol{Q}^{+}-\boldsymbol{Q}^{-}}{\alpha}\right)
        \!\cdot\nabla(\boldsymbol{Q}^{+}+\boldsymbol{Q}^{-})
      =0 .
\end{align}

We now evaluate this relation in the shock limit. Writing the remaining terms in the shock frame removes the explicit time-derivative contribution, and applying the jump substitutions together with the standard contraction identity for dyadic products,
\[
(\mathbf{A}\mathbf{B})_{ij}=A_iB_j,
\qquad
(\mathbf{n}\cdot(\mathbf{A}\mathbf{B}))_j=n_iA_iB_j=(\mathbf{n}\cdot\mathbf{A})B_j,
\]
so that
\[
\mathbf{n}\cdot(\mathbf{A}\mathbf{B})=(\mathbf{n}\cdot\mathbf{A})\,\mathbf{B},
\]
gives
\begin{align}
\label{eq:jump_prime_form}
    \bigg\llbracket
        \left(\frac{Q^{\prime+}_{n}+Q^{\prime-}_{n}}{\alpha}\right)
        (\boldsymbol{Q}^{+}-\boldsymbol{Q}^{-})
    \bigg\rrbracket
    - \left(\frac{Q^{\prime+}_{n}-Q^{\prime-}_{n}}{\alpha}\right)
      \bigg\llbracket \boldsymbol{Q}^{+}+\boldsymbol{Q}^{-} \bigg\rrbracket
    = 0 .
\end{align}

Using the previously derived RH condition for the normal magnetic flux,
equation~\eqref{eq:RH1}, which implies
$\llbracket (Q^{\prime+}_{n}-Q^{\prime-}_{n})/\alpha \rrbracket = 0$,
the prefactor $(Q^{\prime+}_{n}-Q^{\prime-}_{n})/\alpha$ is continuous across the discontinuity and may therefore be factored through the jump operator. Subtracting common terms then yields
\begin{equation}
\label{eq:jump_simplified}
    \bigg\llbracket
        \frac{1}{\alpha}
        \left(Q^{\prime-}_{n}\,\boldsymbol{Q}^{+}
             - Q^{\prime+}_{n}\,\boldsymbol{Q}^{-}\right)
    \bigg\rrbracket
    = 0.
\end{equation}

Finally, we decompose
\[
    \boldsymbol{Q}^{\pm}
    = \boldsymbol{Q}^{\pm}_{t}
      + Q^{\pm}_{n}\mathbf{n},
\]
into tangential and normal parts and add the vanishing term
\[
    u\mathbf{n}\,
    \biggl\llbracket \frac{Q^{\prime+}_{n}-Q^{\prime-}_{n}}{\alpha} \biggr\rrbracket
    = 0
\]
to eliminate the remaining normal-normal contributions. This yields the third Rankine-Hugoniot condition:
\begin{equation}
\label{R-H3}
    \bigg\llbracket
        \frac{
            Q^{\prime-}_{n}\,\boldsymbol{Q}^{\prime+}_{t}
            - Q^{\prime+}_{n}\,\boldsymbol{Q}^{\prime-}_{t}
        }{\alpha}
    \bigg\rrbracket
    = 0 .
\end{equation}
This relation represents the conservation of tangential magnetic flux in the Q-variable formalism.

\subsubsection{Normal momentum condition}

To derive the jump condition associated with the normal component of the momentum, we consider the sum of the two Q-momentum equations~\eqref{eq:momentum} for the $+$ and $-$ components. Adding these equations gives
\begin{align}
    &\frac{D^-}{Dt}\boldsymbol{Q}^+
     +\frac{D^+}{Dt}\boldsymbol{Q}^-
     -\left(\frac{\boldsymbol{Q}^+-\boldsymbol{Q}^-}{4}\right)
      \left(
          \frac{D^-}{Dt}\ln(\rho\alpha^2)
        - \frac{D^+}{Dt}\ln(\rho\alpha^2)
        - \frac{D^-}{Dt}\ln\rho
        + \frac{D^+}{Dt}\ln\rho
      \right) \nonumber\\
    &\quad
     = -2v_s^2\nabla\ln\rho
       -\frac{1}{4}\left(1-\frac{\Delta\alpha^2}{\alpha^2}\right)
        \nabla\left(\boldsymbol{Q}^+-\boldsymbol{Q}^-\right)^2
       +\frac{1}{2}\left(1-\frac{\Delta\alpha^2}{\alpha^2}\right)
        (\boldsymbol{Q}^+-\boldsymbol{Q}^-)^2\nabla\ln\alpha \nonumber\\
    &\quad
      -\frac{1}{2}\frac{\Delta\alpha^2}{\alpha^2}
        (\boldsymbol{Q}^+-\boldsymbol{Q}^-)\cdot\nabla(\boldsymbol{Q}^+-\boldsymbol{Q}^-)
      +\frac{1}{2}\frac{\Delta\alpha^2}{\alpha^2}
        (\boldsymbol{Q}^+-\boldsymbol{Q}^-)\nabla\cdot(\boldsymbol{Q}^+-\boldsymbol{Q}^-).
\end{align}

The terms proportional to $D^\pm\ln\rho/Dt$ cancel once
\[
\ln(\rho\alpha^2)=\ln\rho+\ln\alpha^2,
\]
so that the density contributions enter with opposite signs in the $D^+/Dt$ and $D^-/Dt$ combinations. We then use the solenoidal constraint~\eqref{eq:solenoidal} to write
\begin{equation}
    \frac{D^-}{Dt}\ln\alpha-\frac{D^+}{Dt}\ln\alpha
    = -(\boldsymbol{Q}^+-\boldsymbol{Q}^-)\cdot\nabla\alpha
    = -\nabla\cdot(\boldsymbol{Q}^+-\boldsymbol{Q}^-).
\end{equation}
Using also $v_s^2\nabla\ln\rho = \nabla p/\rho$ and the definition
$\Delta\alpha^2\equiv \alpha^2- 1/(\mu\rho)$, we rewrite the equation as
\begin{align}
    &\frac{D^-}{Dt}\boldsymbol{Q}^+
    +\frac{D^+}{Dt}\boldsymbol{Q}^-
    = -2\frac{\nabla p}{\rho}
      -\frac{1}{4\mu\rho\alpha^2}
        \nabla\left(\boldsymbol{Q}^+-\boldsymbol{Q}^-\right)^2
      +\frac{1}{2\mu\rho\alpha^2}
        (\boldsymbol{Q}^+-\boldsymbol{Q}^-)^2\nabla\ln\alpha \nonumber\\
    &\quad
      -\frac{1}{2}\left(1-\frac{1}{\mu\rho\alpha^2}\right)
        (\boldsymbol{Q}^+-\boldsymbol{Q}^-)\cdot\nabla(\boldsymbol{Q}^+-\boldsymbol{Q}^-)
      -\frac{1}{2\mu\rho\alpha^2}
        (\boldsymbol{Q}^+-\boldsymbol{Q}^-)\nabla\cdot(\boldsymbol{Q}^+-\boldsymbol{Q}^-).
\end{align}

Next, we expand the comoving derivatives and multiply the equation by $2\rho$. After rewriting the logarithmic derivatives of $\alpha$ and collecting similar terms, we obtain
\begin{align}
    &2\rho\frac{\partial}{\partial t}(\boldsymbol{Q}^++\boldsymbol{Q}^-)
     +\rho(\boldsymbol{Q}^++\boldsymbol{Q}^-)\cdot\nabla(\boldsymbol{Q}^++\boldsymbol{Q}^-)
     +4\nabla p \nonumber\\
    &\quad
     +\frac{1}{2\mu\alpha^2}
        \nabla\left(\boldsymbol{Q}^+-\boldsymbol{Q}^-\right)^2
     -\frac{1}{\mu\alpha^3}
        (\boldsymbol{Q}^+-\boldsymbol{Q}^-)^2\nabla\alpha \nonumber\\
    &\quad
     -\frac{1}{\mu\alpha^2}
        (\boldsymbol{Q}^+-\boldsymbol{Q}^-)\cdot\nabla(\boldsymbol{Q}^+-\boldsymbol{Q}^-)
     +\frac{1}{\mu\alpha^2}
        (\boldsymbol{Q}^+-\boldsymbol{Q}^-)\nabla\cdot(\boldsymbol{Q}^+-\boldsymbol{Q}^-)
     = 0.
\end{align}

The last two pairs of terms can be combined using the quotient rule. For the first two terms, we use equation~\eqref{eq:mass_plusminus} to show
\begin{align}
    &2\frac{\partial}{\partial t}\bigl(\rho(\boldsymbol{Q}^++\boldsymbol{Q}^-)\bigr)
      -2(\boldsymbol{Q}^++\boldsymbol{Q}^-)\frac{\partial\rho}{\partial t}
      +\rho(\boldsymbol{Q}^++\boldsymbol{Q}^-)\cdot\nabla(\boldsymbol{Q}^++\boldsymbol{Q}^-) \nonumber\\
    &\quad
      = 2\frac{\partial}{\partial t}\bigl(\rho(\boldsymbol{Q}^++\boldsymbol{Q}^-)\bigr)
       +(\boldsymbol{Q}^++\boldsymbol{Q}^-)\nabla\cdot\bigl(\rho(\boldsymbol{Q}^++\boldsymbol{Q}^-)\bigr)
       +\rho(\boldsymbol{Q}^++\boldsymbol{Q}^-)\cdot\nabla(\boldsymbol{Q}^++\boldsymbol{Q}^-) \nonumber\\
    &\quad
      = 2\frac{\partial}{\partial t}\bigl(\rho(\boldsymbol{Q}^++\boldsymbol{Q}^-)\bigr)
       +\nabla\cdot\bigl(\rho(\boldsymbol{Q}^++\boldsymbol{Q}^-)(\boldsymbol{Q}^++\boldsymbol{Q}^-)\bigr).
\end{align}
Using this, the expression can be written in the form
\begin{align}
    &2\frac{\partial}{\partial t}\bigl(\rho(\boldsymbol{Q}^++\boldsymbol{Q}^-)\bigr)
     +\nabla\cdot\bigl(\rho(\boldsymbol{Q}^++\boldsymbol{Q}^-)(\boldsymbol{Q}^++\boldsymbol{Q}^-)\bigr)
     +4\nabla p \nonumber\\
    &\qquad
     +\frac{1}{2\mu}\nabla\left(\frac{(\boldsymbol{Q}^+-\boldsymbol{Q}^-)^2}{\alpha^2}\right)
     -\frac{1}{\mu}\left(\frac{\boldsymbol{Q}^+-\boldsymbol{Q}^-}{\alpha}\right)
        \cdot\nabla\left(\frac{\boldsymbol{Q}^+-\boldsymbol{Q}^-}{\alpha}\right)
     =0.
\end{align}

This form is now suitable for applying the shock-limit substitutions. Using the results of Section~\ref{sec:Qframework}, we obtain
\begin{align}
    &-2u\llbracket\rho(\boldsymbol{Q}^++\boldsymbol{Q}^-)\rrbracket
     +\mathbf{n}\cdot\llbracket\rho(\boldsymbol{Q}^++\boldsymbol{Q}^-)(\boldsymbol{Q}^++\boldsymbol{Q}^-)\rrbracket
     +4\mathbf{n}\,\llbracket p\rrbracket \nonumber\\
    &\qquad
     +\frac{1}{2\mu}\mathbf{n}\bigg\llbracket
          \frac{(\boldsymbol{Q}^+-\boldsymbol{Q}^-)^2}{\alpha^2}
       \bigg\rrbracket
     -\frac{1}{\mu}\left(\frac{Q^{+}_n-Q^{-}_n}{\alpha}\right)
        \bigg\llbracket \frac{\boldsymbol{Q}^+-\boldsymbol{Q}^-}{\alpha}\bigg\rrbracket
     =0.
\end{align}
Combining the first two terms and transforming to the shock frame yields the total momentum jump condition
\begin{align}
\label{Full momentum}
    &\llbracket\rho(Q^{\prime+}_n+Q^{\prime-}_n)(\boldsymbol{Q}^++\boldsymbol{Q}^-)\rrbracket
     +4\mathbf{n}\,\llbracket p\rrbracket \nonumber\\
    &\qquad
     +\frac{1}{2\mu}\mathbf{n}\bigg\llbracket
         \frac{(\boldsymbol{Q}^+-\boldsymbol{Q}^-)^2}{\alpha^2}
       \bigg\rrbracket
     -\frac{1}{\mu}\left(\frac{Q^{+}_n-Q^{-}_n}{\alpha}\right)
        \bigg\llbracket \frac{\boldsymbol{Q}^+-\boldsymbol{Q}^-}{\alpha}\bigg\rrbracket
     =0.
\end{align}

Equation~\eqref{Full momentum} contains both the normal and tangential momentum balances. To isolate the normal momentum condition, we project along the normal and use the mass-conservation jump condition~\eqref{eq:RH2}. Taking $\mathbf{n}\cdot$ of~\eqref{Full momentum} and adding the identity
\[
  -2u\bigl\llbracket \rho(Q^{\prime+}_n+Q^{\prime-}_n)\bigr\rrbracket = 0,
\]
we obtain
\begin{align}
    &\llbracket\rho(Q^{\prime+}_n+Q^{\prime-}_n)(Q^{\prime+}_n+Q^{\prime-}_n)\rrbracket
     +4\llbracket p\rrbracket \nonumber\\
    &\qquad
     +\frac{1}{2\mu}\bigg\llbracket
          \frac{(\boldsymbol{Q}^+-\boldsymbol{Q}^-)^2}{\alpha^2}
       \bigg\rrbracket
     -\frac{1}{\mu}\left(\frac{Q^{\prime+}_n-Q^{\prime-}_n}{\alpha}\right)
        \bigg\llbracket \frac{Q^{\prime+}_n-Q^{\prime-}_n}{\alpha}\bigg\rrbracket
     =0.
\end{align}

The last term vanishes by virtue of the normal-flux condition~\eqref{eq:RH1}. Writing the third term in terms of normal and tangential components, the normal part vanishes for the same reason, leaving only the tangential contribution. The normal momentum Rankine-Hugoniot condition therefore takes the form
\begin{align}
\label{Normal momentum}
    \bigg\llbracket
        \rho(Q^{\prime+}_n+Q^{\prime-}_n)^2
        + 4 p
        + \frac{1}{2\mu\alpha^2}
          (\boldsymbol{Q}_t^{\prime+}-\boldsymbol{Q}_t^{\prime-})^2
    \bigg\rrbracket
    = 0,
\end{align}
which expresses conservation of the normal momentum flux. 

This relation represents the conservation of normal momentum across the shock in the Q-variable formalism.

\subsubsection{Tangential momentum condition}

To obtain the jump condition associated with the tangential component of the momentum, we start from the full vector momentum relation~\eqref{Full momentum} and subtract its projection along the normal, given by the normal momentum condition~\eqref{Normal momentum}. In other words, we consider
\[
\eqref{Full momentum} - \mathbf{n}\,(\eqref{Normal momentum}).
\]
After transforming to the prime notation of the Q-variables, using the previously derived Rankine-Hugoniot conditions, and cancelling common terms, we obtain
\begin{align}
    &\llbracket \rho(Q^{\prime+}_n+Q^{\prime-}_n)
        (\boldsymbol{Q}^{\prime+}+\boldsymbol{Q}^{\prime-})\rrbracket
      +\frac{1}{2\mu}\mathbf{n}\bigg\llbracket
          \frac{(\boldsymbol{Q}^{\prime+}-\boldsymbol{Q}^{\prime-})^2}{\alpha^2}
        \bigg\rrbracket
      -\frac{1}{\mu}\left(\frac{Q^{\prime+}_n-Q^{\prime-}_n}{\alpha}\right)
        \bigg\llbracket \frac{\boldsymbol{Q}^{\prime+}-\boldsymbol{Q}^{\prime-}}{\alpha}\bigg\rrbracket
      \nonumber\\
    &\qquad
      -\mathbf{n}\bigg\llbracket
          \rho(Q^{\prime+}_n+Q^{\prime-}_n)^2
          +\frac{1}{2\mu\alpha^2}
            (\boldsymbol{Q}_t^{\prime+}-\boldsymbol{Q}_t^{\prime-})^2
        \bigg\rrbracket
      =0.
\end{align}

In the first term, subtracting the normal contribution leaves only the tangential part of the vector, so that the remaining component is
\[
\boldsymbol{Q}^{\prime+}_t + \boldsymbol{Q}^{\prime-}_t.
\]
For the two terms proportional to $\mathbf{n}$, their tangential parts vanish automatically, and their remaining normal parts are zero because the normal-flux Rankine-Hugoniot conditions apply.

In the remaining term, we decompose
\[
  \boldsymbol{Q}^{\prime\pm}
  = \boldsymbol{Q}^{\prime\pm}_t + Q^{\prime\pm}_n\mathbf{n},
\]
and use the normal-flux condition~\eqref{eq:RH1} to eliminate contributions involving
$(Q^{\prime+}_n-Q^{\prime-}_n)/\alpha$ that are continuous across the jump. This leaves the purely tangential vector relation
\begin{align}
    \bigg\llbracket
        \rho(Q^{\prime+}_n+Q^{\prime-}_n)
          (\boldsymbol{Q}^{\prime+}_t+\boldsymbol{Q}^{\prime-}_t)
        - \frac{1}{\mu\alpha^2}
          \left(Q^{\prime+}_n - Q^{\prime-}_n\right)
          \left(\boldsymbol{Q}^{\prime+}_t - \boldsymbol{Q}^{\prime-}_t\right)
    \bigg\rrbracket
    = 0.
\end{align}
This relation expresses the conservation of tangential momentum across the
shock in the Q-variable formalism.

\subsubsection{Energy jump condition}

The final Rankine-Hugoniot condition in classical magnetohydrodynamics follows from the energy conservation law. In \citet{VanDoorsselaere_2024}, the system \eqref{eq:solenoidal}-\eqref{eq:momentum} employs an adiabatic pressure closure rather than a full energy equation, which is sufficient for wave propagation but not for shock jump conditions. Here, however, we require the complete MHD energy conservation law rewritten in the Q-variable formalism.

The classical ideal-MHD energy equation reads
\begin{align}
    &\frac{\partial}{\partial t} \left(\frac{1}{2}\rho V^2  + \frac{p}{\gamma-1} + \frac{1}{2\mu}B^2 \right)
    + \nabla \cdot
    \left[
        \left(
            \frac{1}{2}\rho V^2
            + \frac{\gamma p}{\gamma-1}
            + \frac{1}{\mu}B^2
        \right)\boldsymbol{V}
        - \frac{1}{\mu}(\boldsymbol{V}\cdot\boldsymbol{B})\boldsymbol{B}
    \right]
    = 0 .
\end{align}

Using the relations \eqref{V_and_B}, this equation can be written entirely in terms of the Q-variables as
\begin{align}
&\frac{\partial}{\partial t}
\bigg(
    \frac{1}{8} \rho (\boldsymbol{Q}^+ + \boldsymbol{Q}^-)^2
    + \frac{p}{\gamma-1}
    + \frac{1}{8\mu\alpha^2}
      (\boldsymbol{Q}^+ - \boldsymbol{Q}^-)^2
\bigg)
\nonumber\\
&\quad
+ \nabla \cdot
\bigg[
    \left(
        \frac{1}{8} \rho (\boldsymbol{Q}^+ + \boldsymbol{Q}^-)^2
        + \frac{\gamma p}{\gamma-1}
        + \frac{1}{4\mu\alpha^2}
          (\boldsymbol{Q}^+ - \boldsymbol{Q}^-)^2
    \right)
    \frac{(\boldsymbol{Q}^+ + \boldsymbol{Q}^-)}{2}
\nonumber\\
&\qquad
    - \frac{1}{8\mu\alpha^2}
      \left((\boldsymbol{Q}^{+} + \boldsymbol{Q}^{-})\cdot(\boldsymbol{Q}^{+} - \boldsymbol{Q}^{-})\right)
      (\boldsymbol{Q}^+ - \boldsymbol{Q}^-)
\bigg]
= 0 .
\end{align}

and multiplying by $8$, we obtain
\begin{align}
&2\frac{\partial}{\partial t}
\bigg(
    \frac{1}{2} \rho(\boldsymbol{Q}^+ + \boldsymbol{Q}^-)^2
    + \frac{4p}{\gamma-1}
    + \frac{1}{2\mu\alpha^2}
      (\boldsymbol{Q}^+ - \boldsymbol{Q}^-)^2
\bigg)
\nonumber\\
&\quad
+ \nabla \cdot
\bigg[
    \left(
        \frac{\rho}{2} (\boldsymbol{Q}^+ + \boldsymbol{Q}^-)^2
        + \frac{4\gamma p}{\gamma-1}
        + \frac{1}{\mu\alpha^2}
          (\boldsymbol{Q}^+ - \boldsymbol{Q}^-)^2
    \right)
    (\boldsymbol{Q}^+ + \boldsymbol{Q}^-)
\nonumber\\
&\qquad
    - \frac{1}{\mu\alpha^2}
      \left((\boldsymbol{Q}^+ + \boldsymbol{Q}^-)\cdot(\boldsymbol{Q}^+ - \boldsymbol{Q}^-)\right)
      (\boldsymbol{Q}^+ - \boldsymbol{Q}^-)
\bigg]
= 0 .
\end{align}

This form is suitable for applying the shock-limit substitutions derived in Section~\ref{sec:Qframework}. We therefore obtain
\begin{align}
&-2u\bigg\llbracket
    \frac{\rho}{2} (\boldsymbol{Q}^+ + \boldsymbol{Q}^-)^2
    + \frac{4p}{\gamma-1}
    + \frac{1}{2\mu\alpha^2}
      (\boldsymbol{Q}^+ - \boldsymbol{Q}^-)^2
\bigg\rrbracket
\nonumber\\
&\quad
+ \bigg\llbracket
    \frac{\rho}{2} (\boldsymbol{Q}^+ + \boldsymbol{Q}^-)^2 (Q_n^+ + Q_n^-)
    + \frac{4\gamma p}{\gamma-1} (Q_n^+ + Q_n^-)
\nonumber\\
&\qquad
    + \frac{1}{\mu\alpha^2}
      (\boldsymbol{Q}^+ - \boldsymbol{Q}^-)\!\cdot\!
      (\boldsymbol{Q}^+ - \boldsymbol{Q}^-)(Q_n^+ + Q_n^-)
\nonumber\\
&\qquad
    - \frac{1}{\mu\alpha^2}
      (\boldsymbol{Q}^+ - \boldsymbol{Q}^-)\!\cdot\!
      (\boldsymbol{Q}^+ + \boldsymbol{Q}^-)(Q_n^+ - Q_n^-)
\bigg\rrbracket
= 0 .
\end{align}

To simplify this expression, we add the previously derived normal momentum Rankine-Hugoniot condition, multiplied by $-2u$:
\begin{align}
    -2u\bigg\llbracket
        \rho(Q_n^{\prime+} + Q_n^{\prime-})^2
        + 4p
        + \frac{1}{2\mu\alpha^2}
          (\boldsymbol{Q}_t^{\prime+} - \boldsymbol{Q}_t^{\prime-})^2
    \bigg\rrbracket
    = 0 .
\end{align}

We first consider the pressure-related terms:
\begin{align}
    &-2u\bigg\llbracket \frac{4p}{\gamma-1} \bigg\rrbracket
     -2u\bigg\llbracket 4p \bigg\rrbracket
     +\bigg\llbracket
         \frac{4\gamma p}{\gamma-1} (Q_n^+ + Q_n^-)
     \bigg\rrbracket
    = \bigg\llbracket
        \frac{4\gamma p}{\gamma-1}
        (Q_n^{\prime+} + Q_n^{\prime-})
      \bigg\rrbracket .
\end{align}
The first two terms cancel once the final term is written in shock-frame notation.

Next, we rewrite the quadratic terms by decomposing the dot products into normal and tangential components:
\begin{align}
\label{above}
    &\bigg\llbracket
        \frac{1}{\mu\alpha^2}
        \bigg[
            \bigl((Q_n^+ - Q_n^-)^2
            + (\boldsymbol{Q}_t^+ - \boldsymbol{Q}_t^-)^2\bigr)
            (Q_n^+ + Q_n^-)
\nonumber\\
&\qquad
            - \bigl(
                (\boldsymbol{Q}_t^+ - \boldsymbol{Q}_t^-)\!\cdot\!
                (\boldsymbol{Q}_t^+ + \boldsymbol{Q}_t^-)
                + (Q_n^+ - Q_n^-)(Q_n^+ + Q_n^-)
              \bigr)
              (Q_n^+ - Q_n^-)
        \bigg]
    \bigg\rrbracket .
\end{align}

The first and last terms cancel identically. The remaining term can then be rewritten fully in prime notation.

We also consider the remaining contribution obtained from the time-dependent terms after decomposing $\boldsymbol{Q}$ into tangential and normal components:
\begin{align}
    &-u\bigg\llbracket
        \frac{1}{\mu\alpha^2}
        \Big(
            (Q_n^+ - Q_n^-)^2
            + 2(\boldsymbol{Q}_t^+ - \boldsymbol{Q}_t^-)^2
        \Big)
    \bigg\rrbracket .
\end{align}
Its tangential contribution is used to transform the corresponding term in \eqref{above} to shock-frame notation, while the normal part vanishes by the normal-flux Rankine-Hugoniot condition~\eqref{eq:RH1}.

Collecting all terms, the energy jump condition becomes
\begin{align}
&\bigg\llbracket
        \frac{\rho}{2}
        (\boldsymbol{Q}^+ + \boldsymbol{Q}^-)^2
        (Q_n^{\prime+} + Q_n^{\prime-})
        + \frac{4\gamma p}{\gamma-1}
          (Q_n^{\prime+} + Q_n^{\prime-})
        + \frac{1}{\mu\alpha^2}
          (\boldsymbol{Q}_t^{\prime+} - \boldsymbol{Q}_t^{\prime-})^2
          (Q_n^{\prime+} + Q_n^{\prime-})
\nonumber\\
&\qquad
        -\frac{1}{\mu\alpha^2}
        (\boldsymbol Q^{\prime+}_t-\boldsymbol Q^{\prime-}_t)\cdot(\boldsymbol{Q}^{\prime+}_t+\boldsymbol{Q}^{\prime-}_t)
        (Q^{+}_n-Q^{-}_n)
        - 2u\rho (Q_n^{\prime+} + Q_n^{\prime-})^2
\bigg\rrbracket
= 0 .
\end{align}

We now expand the kinetic-energy term into tangential and normal parts and combine it with the last term:
\begin{align}
    \frac{\rho}{2}(Q_n^{\prime+} + Q_n^{\prime-})
    \left[
        (\boldsymbol{Q}_t^+ + \boldsymbol{Q}_t^-)^2
        + (Q_n^+ + Q_n^-)^2
        - 4u(Q_n^{\prime+} + Q_n^{\prime-})
    \right].
\end{align}
Using the definition of the prime variables and expanding the square of the normal sum yields
\begin{align}
    \frac{\rho}{2}(Q_n^{\prime+} + Q_n^{\prime-})
    \big[
        (\boldsymbol{Q}_t^+ + \boldsymbol{Q}_t^-)^2
        + (Q_n^{\prime+} + Q_n^{\prime-})^2
        + 4u^2
    \big].
\end{align}

The $4u^2$ term vanishes due to the mass-conservation jump condition~\eqref{eq:RH2}. The final Rankine-Hugoniot condition:
\begin{align}
\label{Energy}
&\bigg\llbracket
        \frac{\rho}{2}(Q_n^{\prime+} + Q_n^{\prime-})
        \left[
            (\boldsymbol{Q}_t^{\prime+} + \boldsymbol{Q}_t^{\prime-})^2
            + (Q_n^{\prime+} + Q_n^{\prime-})^2
        \right]
        + \frac{4\gamma p}{\gamma-1}
          (Q_n^{\prime+} + Q_n^{\prime-})
\nonumber\\
&\qquad
        + \frac{2}{\mu\alpha^2}
        \left(
            Q_n^{\prime-}\boldsymbol{Q}_t^{\prime+}
            - Q_n^{\prime+}\boldsymbol{Q}_t^{\prime-}
        \right)
        \!\cdot\!
        (\boldsymbol{Q}_t^{\prime+} - \boldsymbol{Q}_t^{\prime-})
\bigg\rrbracket
= 0 ,
\end{align}
which expresses the conservation of energy across the shock in the Q-variable formalism.

\subsection{Final set of Rankine-Hugoniot conditions in Q-variables}
\label{sec:RH_summary}

We now collect the full set of Rankine-Hugoniot conditions derived in terms of the Q-variables. These relations describe the jump conditions across a steady MHD shock in the Q-variable formalism.

\begin{align}
    &\bigg\llbracket \frac{Q^{\prime+}_n - Q^{\prime-}_n}{\alpha} \bigg\rrbracket = 0,
    \label{Normal flux_sys} \\[4pt]
    &\bigg\llbracket \rho\,(Q^{\prime+}_n + Q^{\prime-}_n) \bigg\rrbracket = 0,
    \label{Mass_sys} \\[4pt]
    &\bigg\llbracket
        \frac{
            Q^{\prime-}_n \boldsymbol{Q}^{\prime+}_t
            - Q^{\prime+}_n \boldsymbol{Q}^{\prime-}_t
        }{\alpha}
    \bigg\rrbracket = 0,
    \label{Tangential flux_sys} \\[6pt]
    &    \bigg\llbracket
        \rho(Q^{\prime+}_n+Q^{\prime-}_n)^2
        + 4 p
        + \frac{1}{2\mu\alpha^2}
          (\boldsymbol{Q}_t^{\prime+}-\boldsymbol{Q}_t^{\prime-})^2
    \bigg\rrbracket
    = 0,
    \label{Normal momentum_sys} \\[6pt]
    &    \bigg\llbracket
        \rho(Q^{\prime+}_n+Q^{\prime-}_n)
          (\boldsymbol{Q}^{\prime+}_t+\boldsymbol{Q}^{\prime-}_t)
        - \frac{1}{\mu\alpha^2}
          \left(Q^{\prime+}_n - Q^{\prime-}_n\right)
          \left(\boldsymbol{Q}^{\prime+}_t - \boldsymbol{Q}^{\prime-}_t\right)
    \bigg\rrbracket=0,
    \label{Tangential momentum_sys} \\[6pt]
    &\bigg\llbracket
        \frac{\rho}{2}(Q_n^{\prime+} + Q_n^{\prime-})
        \left[
            (\boldsymbol{Q}_t^{\prime+} + \boldsymbol{Q}_t^{\prime-})^2
            + (Q_n^{\prime+} + Q_n^{\prime-})^2
        \right]
        + \frac{4\gamma p}{\gamma-1}
          (Q_n^{\prime+} + Q_n^{\prime-})
\nonumber\\
&\qquad
        + \frac{2}{\mu\alpha^2}
        \left(
            Q_n^{\prime-}\boldsymbol{Q}_t^{\prime+}
            - Q_n^{\prime+}\boldsymbol{Q}_t^{\prime-}
        \right)
        \!\cdot\!
        (\boldsymbol{Q}_t^{\prime+} - \boldsymbol{Q}_t^{\prime-})
\bigg\rrbracket=0.
    \label{Energy_sys}
\end{align}

Equations~\eqref{Normal flux_sys}--\eqref{Energy_sys} represent, respectively, the normal magnetic-flux condition, mass conservation, tangential magnetic-flux condition, normal momentum conservation, tangential momentum conservation, and energy conservation across the shock, all expressed in the Q-variable formalism.

\subsection{Recovery of the classical Rankine-Hugoniot conditions from Q-variables}

We now verify that the Rankine-Hugoniot conditions derived in the Q-variable formalism reduce to the classical MHD jump conditions when expressed in the standard variables $(\rho,\boldsymbol{V},p,\boldsymbol{B})$. The equivalence proof presented below is performed exclusively using the conservative forms ~\eqref{Normal flux_sys}--\eqref{Energy_sys}. The alternative relations of Section~\ref{Alternative} are derived only after equivalence has been established and do not constitute independent Rankine-Hugoniot conditions.

We begin with the normal flux condition~\eqref{Normal flux_sys}. Using the definition of the Q-variables, we introduce the velocity in the shock-comoving frame,
\[
    V_n' = V_n - u ,
\]
so that
\[
    Q_n^{\prime\pm} = V_n' \pm \alpha B_n,
    \qquad
    \boldsymbol{Q}_t^{\prime\pm}
    = \boldsymbol{V}_t' \pm \alpha \boldsymbol{B}_t .
\]
Substituting into~\eqref{Normal flux_sys} yields
\begin{align}
    \bigg\llbracket
        \frac{V_n' + \alpha B_n - V_n' + \alpha B_n}{\alpha}
    \bigg\rrbracket
    = \bigg\llbracket B_n \bigg\rrbracket
    = 0 .
\end{align}
This immediately recovers the classical jump condition expressing continuity of the normal magnetic field across the shock.

Next, we consider the mass conservation condition~\eqref{Mass_sys}:
\begin{align}
    \bigg\llbracket
        \rho \left( V_n' + \alpha B_n + V_n' - \alpha B_n \right)
    \bigg\rrbracket
    = \bigg\llbracket \rho V_n' \bigg\rrbracket
    = 0 .
\end{align}
This is precisely the classical mass-flux jump condition.

We now consider the tangential flux condition~\eqref{Tangential flux_sys}. Substituting the definitions of the Q-variables gives
\begin{align}
    \bigg\llbracket
        \frac{
            (V_n' - \alpha B_n)(\boldsymbol{V}_t' + \alpha \boldsymbol{B}_t)
            - (V_n' + \alpha B_n)(\boldsymbol{V}_t' - \alpha \boldsymbol{B}_t)
        }{\alpha}
    \bigg\rrbracket
    = \bigg\llbracket
        V_n' \boldsymbol{B}_t - B_n \boldsymbol{V}_t'
    \bigg\rrbracket
    = 0 .
\end{align}
This is exactly the classical tangential magnetic-flux jump condition.

Next, we consider the normal momentum condition~\eqref{Normal momentum_sys}. Using the identities \eqref{V_and_B} and dividing the resulting expression by $4$, we obtain
\begin{align}
    \bigg\llbracket
        \rho V_n'^2
        + p
        + \frac{1}{2\mu} \boldsymbol{B}_t^2
    \bigg\rrbracket
    = 0 .
\end{align}
This is the classical normal momentum Rankine-Hugoniot condition.

We proceed with the tangential momentum condition~\eqref{Tangential momentum_sys}. Applying the same substitutions and dividing by $4$, we find
\begin{align}
    \bigg\llbracket
        \rho V_n' \boldsymbol{V}_t
        - \frac{1}{\mu} \boldsymbol{B}_t B_n
    \bigg\rrbracket
    = 0 .
\end{align}
This coincides with the classical tangential momentum jump condition.

Finally, we consider the energy jump condition~\eqref{Energy_sys}. Expressed in standard variables, this condition takes the form
\begin{align}
    &\bigg\llbracket
       4\rho V_n'
        \left(
            V_n'^2 + \boldsymbol{V}_t^2
        \right)
        + 8 V_n' \frac{\gamma p}{\gamma - 1} 
        \nonumber\\
    &\qquad
        + \frac{4}{\mu \alpha}
        \bigg[
            (V_n' - \alpha B_n)(\boldsymbol{V}_t + \alpha \boldsymbol{B}_t)
            - (V_n' + \alpha B_n)(\boldsymbol{V}_t - \alpha \boldsymbol{B}_t)
        \bigg]
        \!\cdot\! \boldsymbol{B}_t
    \bigg\rrbracket
    = 0 .
\end{align}
Simplifying this expression and dividing it by 8, we obtain
\begin{align}
    \bigg\llbracket
        \frac{1}{2}\rho V_n'
        \left(
            V_n'^2 + \boldsymbol{V}_t^2
        \right)
        + V_n' \frac{\gamma p}{\gamma - 1}
        + \frac{1}{\mu} V_n' \boldsymbol{B}_t^2
        - \frac{1}{\mu} B_n \boldsymbol{V}_t \cdot \boldsymbol{B}_t
    \bigg\rrbracket
    = 0 .
\end{align}
This is precisely the classical MHD energy Rankine-Hugoniot condition.

Altogether, these results demonstrate that the Q-variable formalism recovers the full set of classical MHD jump conditions. This confirms the robustness of the Q-variable formalism and its applicability to shock physics within ideal magnetohydrodynamics.

\subsection{Alternative representation of the Rankine-Hugoniot conditions}
\label{Alternative}

The conservative Rankine-Hugoniot conditions derived in the previous sections constitute the primary jump relations used to establish exact equivalence with the classical MHD system. However, in classical shock theory it is often convenient to rewrite the momentum and energy fluxes in forms that make particular physical parameters explicit \citep{Landau_Lifshitz_1984, Goedbloed_Keppens_Poedts_2019}.

Importantly, this procedure does not involve dividing a Rankine-Hugoniot condition by a discontinuous density. Instead, the conservative fluxes are first evaluated independently in the upstream and downstream states and are then rewritten algebraically using the local plasma parameters. Since $\rho>0$ in ideal MHD, such transformations are well defined and remain exactly equivalent to the conservative jump conditions.

Within the Q-variable formalism, it is convenient to use the relation
\begin{equation}
\Delta\alpha^2 =\alpha^2
-\frac{1}{\mu\rho},
\end{equation}
which implies
\begin{equation}
\frac{1}{\mu\rho\alpha^2}=
1-\frac{\Delta\alpha^2}{\alpha^2}.
\end{equation}

Here $\Delta\alpha^2$ denotes the algebraic quantity introduced in the Q-variable formalism; it measures the deviation of the selected wave parameter from its Alfvénic value and is not a differential operator. Since $\alpha$ is determined locally by the plasma state and the selected wave branch, this identity may be applied independently on each side of the discontinuity. The resulting expressions therefore represent an alternative form of the conservative Rankine-Hugoniot system rather than an additional set of jump conditions.

Using this identity, the normal momentum condition \eqref{Normal momentum} may be written as
\begin{align}
\bigg\llbracket
(Q_n^{\prime+}+Q_n^{\prime-})^2
+
\frac{4p}{\rho}
+
\frac12
\left(
1-\frac{\Delta\alpha^2}{\alpha^2}
\right)
(\boldsymbol{Q}_t^{\prime+}-\boldsymbol{Q}_t^{\prime-})^2
\bigg\rrbracket
=0.
\label{Normal_momentum_specific}
\end{align}

Similarly, the tangential momentum condition becomes
\begin{align}
\bigg\llbracket
(Q_n^{\prime+}+Q_n^{\prime-})
(\boldsymbol{Q}_t^{\prime+}+\boldsymbol{Q}_t^{\prime-})-
\left(
1-\frac{\Delta\alpha^2}{\alpha^2}
\right)
(Q_n^{\prime+}-Q_n^{\prime-})
(\boldsymbol{Q}_t^{\prime+}-\boldsymbol{Q}_t^{\prime-})
\bigg\rrbracket
=0,
\label{Tangential_momentum_specific}
\end{align}

or, after expanding the products and rearranging terms,
\begin{align}
\bigg\llbracket
2Q_n^{\prime-}\boldsymbol{Q}_t^{\prime+}
+
2Q_n^{\prime+}\boldsymbol{Q}_t^{\prime-}
+
\frac{\Delta\alpha^2}{\alpha^2}
\left(
Q_n^{\prime+}-Q_n^{\prime-}
\right)
\left(
\boldsymbol{Q}_t^{\prime+}-\boldsymbol{Q}_t^{\prime-}
\right)
\bigg\rrbracket
=0.
\label{Tangential_momentum_symmetric}
\end{align}

Proceeding in the same manner, the energy condition may be written as

\begin{align}
&\bigg\llbracket
        \frac{1}{2}
        (Q^{\prime+}_n + Q^{\prime-}_n)
        \Big[
            (\boldsymbol{Q}^{\prime+}_t + \boldsymbol{Q}^{\prime-}_t)^2
            + (Q^{\prime+}_n + Q^{\prime-}_n)^2
        \Big]
        + \frac{4\gamma p}{\rho(\gamma-1)}
          (Q^{\prime+}_n + Q^{\prime-}_n)
        \nonumber\\
    &\qquad\qquad
        + 2\left(1 - \frac{\Delta \alpha^2}{\alpha^2}\right)
        \left(
            Q^{\prime-}_n \boldsymbol{Q}^{\prime+}_t
            - Q^{\prime+}_n \boldsymbol{Q}^{\prime-}_t
        \right)
        \!\cdot\!
        (\boldsymbol{Q}^{\prime+}_t - \boldsymbol{Q}^{\prime-}_t)
    \bigg\rrbracket = 0.
\label{Energy_specific}
\end{align}

These expressions are algebraically equivalent to the conservative Rankine-Hugoniot relations derived above. Their advantage is that the dependence on the local wave parameter $\alpha$ becomes explicit, which can be useful when analysing branch-dependent limits and characteristic interpretations of MHD discontinuities. Nevertheless, the conservative forms derived in Section~III remain the fundamental Rankine-Hugoniot relations and are used throughout this work when establishing equivalence with the classical MHD jump conditions.

\section{Wave-aligned interpretation of discontinuities}
\label{sec:interpretation}

The Rankine-Hugoniot conditions derived in the previous section establish that the Q-variable formalism provides a complete and self-consistent description of MHD discontinuities. In addition to reproducing the classical jump conditions exactly, the Q-variable formulation offers a complementary interpretation in which the characteristic contributions of the flow are made explicit across discontinuities. 

It is important to clarify the meaning of wave alignment in the compressible case. For Alfvénic perturbations, the Q-variables reduce to the Elsässer variables and provide the familiar separation of oppositely propagating Alfvénic contributions. For magnetosonic perturbations, the situation is more structured: fast and slow waves involve density and pressure variations in addition to velocity and magnetic-field perturbations. In the Q-variable formalism, this thermodynamic dependence enters through the branch-dependent parameter $\alpha$, which is determined by the corresponding dispersion relation and therefore depends on the local plasma state. Thus, the Q-variables should not be understood as removing the density and pressure degrees of freedom from compressible MHD. Rather, they provide a wave-aligned representation of the velocity-magnetic-field combinations associated with a chosen characteristic branch, while the thermodynamic variables remain coupled through the Rankine-Hugoniot conditions.

Rather than analysing discontinuities solely in terms of bulk plasma variables, the Q-variables reorganise the dynamics into combinations aligned with propagation directions determined by the parameter $\alpha$. This representation makes the directional contributions associated with oppositely propagating characteristics explicit across a discontinuity, which are otherwise mixed in the classical variables.

To illustrate this wave-aligned interpretation, we consider a simple one-dimensional Riemann-type example.

\subsection{Wave-aligned interpretation: Riemann-type example}

We consider a schematic one-dimensional discontinuity and compare its representation in classical MHD variables with that in Q-variables. The purpose of this example is not to perform a full solution of the Riemann problem, but rather to demonstrate how the Q-formalism represents the velocity-magnetic-field variables in a form aligned with propagation directions across a discontinuity.

\begin{figure}[ht]
    \centering
    \includegraphics[width=1\textwidth]{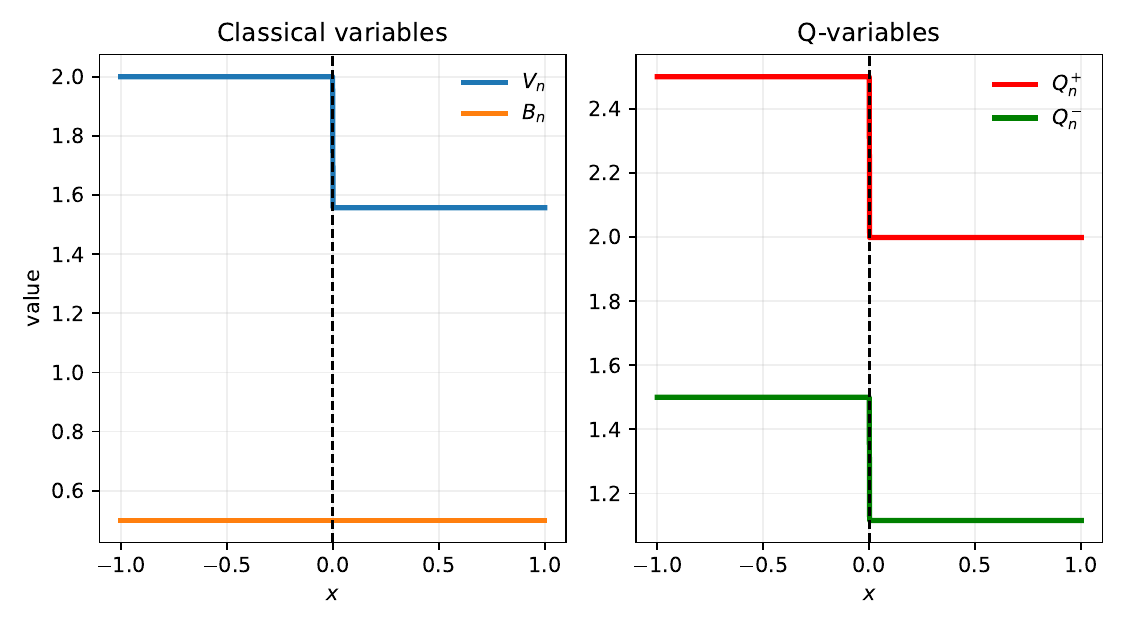}
    \caption{Comparison of a one-dimensional discontinuity in classical MHD variables (left panel) and Q-variables (right panel). While the normal magnetic field $B_n$ remains continuous across the discontinuity, the Q-variables display the wave-aligned structure and contributions associated with oppositely directed characteristic branches.}
    \label{fig:riemann_Q_vs_classical}
\end{figure}

Figure~\ref{fig:riemann_Q_vs_classical} shows a one-dimensional setup with an initial discontinuity located at $x=0$. The left and right states satisfy the Rankine-Hugoniot conditions derived in the previous sections.

In the classical representation (left panel), the normal velocity $V_n$ and magnetic field $B_n$ are shown. As required by the Rankine-Hugoniot conditions, $B_n$ remains continuous across the discontinuity, while $V_n$ exhibits a jump. In these variables, the underlying wave content of the discontinuity is not explicitly separated.

In the Q-variable representation (right panel), the same discontinuity is expressed in terms of $Q_n^{+}$ and $Q_n^{-}$. These variables correspond to oppositely directed propagation branches associated with the chosen parameter $\alpha$. The discontinuity can therefore be represented in terms of contributions associated with oppositely directed characteristic branches, consistent with the interpretation of shocks as discontinuities associated with characteristic fields in hyperbolic systems \citep{Jeffrey_Taniuti_2000}.

Here, we adopt the Elsässer choice $\alpha = 1/\sqrt{\mu\rho}$, evaluated separately on each side of the discontinuity. Even in this simple example, the Q-variable representation displays contributions from both propagation directions, despite the continuity of $B_n$ in the classical variables.

This illustrates a key advantage of the Q-variable formalism: even when the classical variables suggest a simple structure (for example, continuity of the normal magnetic field), the Q-variable representation makes the directional characteristic contributions explicit. This provides a natural framework for analysing the directional structure of wave-discontinuity interactions.

In particular, when $\alpha$ is chosen to match a characteristic speed, the Q-variables align with individual propagation branches. This property is especially useful in situations where one-directional propagation dominates, such as wave transport in stratified or open magnetic structures.

To make this interpretation more concrete, we first consider a standard rotational discontinuity and then specialise to the case of an Alfvénic perturbation associated with this discontinuity.

\subsection{Illustrative example: an Alfvénic perturbation at a rotational discontinuity}

To illustrate more concretely how the Q-variable formulation provides
additional physical insight, we consider a rotational discontinuity and then
specialise to the case in which the relevant wave branch is Alfvénic. The aim
of this example is not to provide a complete classification of MHD
discontinuities in Q-variables. A systematic classification of contact,
tangential, and rotational discontinuities, as well as shock families, will be
addressed in future work. Here, we use the rotational discontinuity only as a
simple example showing how the Q-variable representation makes the directional
wave content of the discontinuity explicit.

A rotational discontinuity is non-compressive. We therefore impose
\[
\llbracket \rho \rrbracket = 0,
\qquad
\llbracket p \rrbracket = 0.
\]
Using the Q-variable Rankine-Hugoniot conditions derived above, the absence of
compression implies continuity of the normal velocity. In Q-variables this may
be written as
\[
\llbracket Q^{\prime+}_n + Q^{\prime-}_n \rrbracket = 0.
\]

The tangential part of the discontinuity contains the rotational information.
For a rotational discontinuity, the reduced Q-variable jump conditions may then
be written as
\begin{align}
    &\llbracket \rho \rrbracket = 0,
    \label{RD_rho} \\[4pt]
    &\llbracket p \rrbracket = 0,
    \label{RD_p} \\[4pt]
    &\llbracket Q^{\prime+}_n + Q^{\prime-}_n \rrbracket = 0,
    \label{RD_normal_velocity} \\[4pt]
    &\bigg\llbracket
        \frac{Q^{\prime+}_n - Q^{\prime-}_n}{\alpha}
    \bigg\rrbracket = 0,
    \label{RD_normal_flux} \\[6pt]
    &\bigg\llbracket
        \boldsymbol{Q}^{\prime+}_t
        + \boldsymbol{Q}^{\prime-}_t
    \bigg\rrbracket
    =
    \pm \frac{1}{\sqrt{\mu\rho}}
    \bigg\llbracket
        \frac{
        \boldsymbol{Q}^{\prime+}_t
        - \boldsymbol{Q}^{\prime-}_t
        }{\alpha}
    \bigg\rrbracket,
    \label{RD_tangential_correlation} \\[6pt]
    &\bigg\llbracket
        \frac{
        \left(
        \boldsymbol{Q}^{\prime+}_t
        - \boldsymbol{Q}^{\prime-}_t
        \right)^2
        }{\alpha^2}
    \bigg\rrbracket = 0.
    \label{RD_quadratic_invariant}
\end{align}
These relations show that, in the Q-variable representation, the rotational
discontinuity is described by a redistribution of tangential wave content
between the two directional components, while the non-compressive character of
the discontinuity is retained through the continuity of density, pressure, and
normal velocity. In particular, the last relation expresses the preservation of
the tangential magnetic-field magnitude, written in Q-variable form.

We now consider the simple case in which an Alfvénic perturbation passes through the rotational discontinuity. In this case the wave parameter appearing in the
Q-variable Rankine-Hugoniot conditions is chosen as
\[
\alpha_A = \frac{1}{\sqrt{\mu\rho}}.
\]
This choice is not imposed by the Rankine-Hugoniot conditions themselves.
Rather, it corresponds to selecting the Alfvén branch of the wave content that
is being represented by the Q-variables. Since the rotational discontinuity is
non-compressive, the density is continuous and therefore
\[
\llbracket \alpha_A \rrbracket = 0.
\]
The normal Q-variable conditions then reduce to
\[
\llbracket Q^{\prime+}_n \rrbracket = 0,
\qquad
\llbracket Q^{\prime-}_n \rrbracket = 0.
\]
Thus, for the Alfvénic branch, the normal wave structure is unchanged across
the discontinuity.

The tangential conditions also simplify. Since $\alpha=\alpha_A$, the
tangential relation becomes
\[
\bigg\llbracket
\boldsymbol{Q}^{\prime+}_t
+
\boldsymbol{Q}^{\prime-}_t
\bigg\rrbracket
=
\pm
\bigg\llbracket
\boldsymbol{Q}^{\prime+}_t
-
\boldsymbol{Q}^{\prime-}_t
\bigg\rrbracket .
\]
Consequently, for the plus sign one obtains
\[
\llbracket \boldsymbol{Q}^{\prime-}_t \rrbracket
=
\boldsymbol{0},
\]
whereas for the minus sign one obtains
\[
\llbracket \boldsymbol{Q}^{\prime+}_t \rrbracket
=
\boldsymbol{0}.
\]
Therefore, only one directional Alfvénic component carries the tangential
change, while the other passes through the discontinuity unchanged. The
quadratic invariant becomes
\[
\bigg\llbracket
\left(
\boldsymbol{Q}^{\prime+}_t
-
\boldsymbol{Q}^{\prime-}_t
\right)^2
\bigg\rrbracket = 0,
\]
which shows that the overall tangential amplitude is preserved.

This example demonstrates the physical advantage of the Q-variable
representation. In the classical variables, the rotational discontinuity is
described by correlated changes of the tangential velocity and magnetic field.
In the Q-variables, the same information becomes a statement about directional
Alfvénic content: one tangential wave component is transmitted unchanged, while
the other is modified so that the structure is rotated but not amplified. The
parameter $\alpha$ therefore acts as the branch-dependent wave parameter of
the perturbation crossing the discontinuity, while the Rankine-Hugoniot
conditions determine how that wave content is constrained by the discontinuity.
This makes the physical interpretation more direct than in the classical
velocity-magnetic-field variables alone. A more systematic treatment of
discontinuity classification in Q-variables, including different discontinuity
families and the interaction of different wave branches with them, will be
addressed in future work.

\subsection{One-directional limits and characteristic content}

A key motivation for introducing the Q-variables is that, for an appropriate
choice of the parameter $\alpha$, they provide a decomposition of the MHD
dynamics aligned with characteristic propagation directions. In this
representation, the $+$ and $-$ fields correspond to oppositely directed
characteristic contributions of the system.

When a discontinuity is considered, these contributions may be projected onto
the normal direction, allowing one to distinguish incoming and outgoing
characteristic content in the shock frame. This provides a natural framework
for analysing directional properties of the solution.

In this section, we consider the limiting case in which the dynamics is
one-directional in the characteristic sense, meaning that only a single characteristic branch is present on either side of the discontinuity.
Importantly, this notion does not refer to the geometry of the discontinuity
itself, but rather to a restriction on the plasma state in which either the
$+$ or the $-$ contribution vanishes.

Imposing $\boldsymbol{Q}^{\prime\pm} = \boldsymbol{0}$ implies
\[
Q^{\prime\pm}_n = 0,
\qquad
\boldsymbol{Q}^{\prime\pm}_t = \boldsymbol{0},
\]
so that both normal and tangential components of the suppressed branch vanish
identically. The dynamics is then entirely carried by the remaining
characteristic branch.

Under either of the single-branch constraints,
$\boldsymbol{Q}^{\prime-} = \boldsymbol{0}$ or
$\boldsymbol{Q}^{\prime+} = \boldsymbol{0}$, all flux contributions involving the suppressed branch vanish identically.
The full Rankine-Hugoniot system, therefore, reduces to
\begin{align}
    &\bigg\llbracket \frac{Q^{\prime\pm}_n}{\alpha} \bigg\rrbracket = 0,
    \label{Normal flux_UD} \\[4pt]
    &\bigg\llbracket \rho Q^{\prime\pm}_n \bigg\rrbracket = 0,
    \label{Mass_UD} \\[4pt]
    &\bigg\llbracket
        Q^{\prime\pm2}_n
        + \frac{4p}{\rho}
        + \frac{1}{2}
          \left(1 - \frac{\Delta \alpha^2}{\alpha^2}\right)
          \boldsymbol{Q}^{\prime\pm2}_t
    \bigg\rrbracket = 0,
    \label{Normal momentum_UD} \\[6pt]
    &\bigg\llbracket
        \frac{\Delta \alpha^2}{\alpha^2}
        Q^{\prime\pm}_n \boldsymbol{Q}^{\prime\pm}_t
    \bigg\rrbracket = 0,
    \label{Tangential momentum_UD} \\[6pt]
    &\bigg\llbracket
        \frac{1}{2}Q^{\prime\pm}_n
        \left(\boldsymbol{Q}^{\prime\pm2}_t + Q^{\prime\pm2}_n\right)
        + Q^{\prime\pm}_n \frac{4\gamma p}{\rho(\gamma-1)}
    \bigg\rrbracket = 0.
    \label{Energy_UD}
\end{align}

In this limit, the tangential flux condition~\eqref{Tangential flux_sys} is
identically satisfied, reflecting the reduction in the number of independent
variables. The remaining relations form a closed and self-consistent set of
jump conditions for one-directional dynamics.

It is important to emphasise that the Rankine-Hugoniot system itself does not
enforce one-directionality. Rather, the single-branch assumption represents a
restriction on the admissible state space. Once imposed, however, the
Q-variable formulation yields a consistent reduced system.

This interpretation is illustrated in
Fig.~\ref{fig:riemann_Q_vs_classical}. While the normal magnetic field remains
continuous in the classical representation, both $Q_n^{+}$ and $Q_n^{-}$
exhibit finite jumps, indicating the presence of contributions from both
characteristic branches. The one-directional limits correspond to selecting only one of these branches, making the characteristic content of the discontinuity explicit.

\subsection{Wave-branch $\alpha$-relations across discontinuities}
\label{sec:alpha_relations}

Beyond the classification of discontinuities and the identification of their directional characteristic content, the Q-variable formulation provides a natural framework for relating branch-dependent wave parameters to the underlying plasma state across a discontinuity. In this formulation, the Rankine-Hugoniot conditions do not prescribe which wave modes may interact with a shock. Rather, they constrain how the parameters characterising a given wave are modified as a consequence of the change in the background plasma.

A key quantity in this description is the parameter $\alpha$, which enters the definition of the Q-variables and is determined by the dispersion relation of the wave mode. For a given disturbance, $\alpha$ depends on the local plasma properties as well as on the wave frequency and wavevector, and therefore does not represent an independent shock variable. Instead, the Rankine-Hugoniot conditions relate the upstream and downstream values of $\alpha$ indirectly through the corresponding changes in density, pressure, and magnetic field.

For Alfv\'en waves, the characteristic choice $\alpha^2 = 1/(\mu \rho)$ implies that $\alpha$ depends only on the plasma density. The jump conditions, therefore, lead to the relation
\begin{equation}
    \frac{\alpha_1}{\alpha_2} = \sqrt{\frac{\rho_2}{\rho_1}},
\end{equation}
which expresses how the characteristic parameter associated with an Alfv\'enic disturbance changes across the discontinuity as a direct consequence of the density jump. This relation does not impose an additional constraint on the discontinuity itself, but instead reflects how a specific wave mode adapts to the modified plasma environment.

For magnetosonic waves, including both fast and slow branches, the parameter $\alpha$ is determined by the corresponding dispersion relation,
\begin{equation}
    \alpha^2 =
    \frac{(V_{A0}^2 + v_{s0}^2)
    \pm \sqrt{(V_{A0}^2 + v_{s0}^2)^2
    - 4 V_{A0}^2 v_{s0}^2 \cos^2\theta}}
    {2k_z^2B_0^2},
\end{equation}
where $V_{A0}$ and $v_{s0}$ denote the Alfv\'en and sound speeds, $\theta$ is the angle between the wavevector and the magnetic field, and the $\pm$ sign corresponds to the fast and slow branches. Variations of the plasma parameters across a discontinuity therefore induce corresponding changes in $\alpha$, reflecting the modification of the local wave propagation properties.

More generally, for a wave propagating in a moving plasma, $\alpha$ can be expressed as
\begin{equation}
    \alpha =
    \frac{\omega - \bm{k}\!\cdot\!\boldsymbol{V}_0}
         {\bm{k}\!\cdot\!\boldsymbol{B}_0}.
\end{equation}
The relation between $\alpha_1$ and $\alpha_2$, together with the corresponding wavevectors and background fields, then characterises how the effective phase speed associated with the chosen wave branch changes across the discontinuity:
\begin{equation}
    \frac{\alpha_1\,(\bm{k}_1\!\cdot\!\boldsymbol{B}_1)}
         {\alpha_2\,(\bm{k}_2\!\cdot\!\boldsymbol{B}_2)}
    =
    \frac{\omega_1 - \bm{k}_1\!\cdot\!\boldsymbol{V}_1}
         {\omega_2 - \bm{k}_2\!\cdot\!\boldsymbol{V}_2}
    \;\equiv\;
    \frac{c_{\alpha,1}}{c_{\alpha,2}}.
\end{equation}
This relation does not constitute a wave-transmission law in the strict sense, as the Rankine-Hugoniot conditions describe nonlinear state matching rather than linear wave propagation. However, it provides a useful characterisation of how the wave-associated parameter $\alpha$ responds to changes in the plasma state across the discontinuity.

From this perspective, the Q-variable formulation offers a direct link between wave properties and discontinuity structure. Changes in $\alpha$ reflect the plasma jump conditions and therefore provide information about the nature of the discontinuity. In observational or numerical settings, measurements of wave propagation properties on either side of a discontinuity may thus be used to constrain the associated plasma parameters and to characterise aspects of the discontinuity structure. While such an inversion is generally not unique, the Q-variable framework provides a physically transparent basis for connecting wave diagnostics with shock and discontinuity physics.

Rather than introducing additional constraints on discontinuities, the Q-variable approach reorganises the Rankine-Hugoniot theory in a form that makes its wave-related content explicit. This highlights the potential of the formalism as a diagnostic tool for studying wave-discontinuity interactions in magnetised plasmas.

\section{Conclusions}
\label{sec:conclusions}

In this work, we have derived the complete set of Rankine-Hugoniot
conditions for ideal magnetohydrodynamics expressed entirely in terms of
the Q-variables. Starting from the ideal MHD system, we reformulated the
conservation laws of mass, momentum, magnetic flux, and energy within the
Q-variable formalism and constructed the corresponding jump conditions
across a planar discontinuity in the shock frame. By explicit
transformation back to the standard $(\rho,\boldsymbol{V},p,\boldsymbol{B})$
variables, we have shown that the resulting relations recover the
classical Rankine-Hugoniot conditions exactly.

Beyond this equivalence, the Q-variable formalism provides an alternative
representation of MHD discontinuities in which the dynamics is organised
according to characteristic propagation directions. For a given choice of
the parameter $\alpha$, the variables $\boldsymbol{Q}^{\pm}$ correspond to
combinations of velocity and magnetic-field fluctuations aligned with
oppositely directed characteristic branches. In this representation, the
jump conditions can be interpreted directly in terms of directional
contributions, rather than solely in terms of bulk conserved quantities.
The illustrative rotational-discontinuity example shows how this representation turns
correlated jumps in tangential velocity and magnetic field into a direct statement about the active
Alfvénic branch.

This characteristic alignment offers a useful perspective for analysing
discontinuities. In particular, it enables a direct identification of how different directional components behave across an interface, without requiring an additional decomposition of the classical velocity and magnetic-field variables. This makes the directional characteristic content of the discontinuity more explicit and provides a natural framework for analysing wave-discontinuity interactions in a branch-dependent representation. The formalism also naturally accommodates one-directional limits, in which only a single characteristic branch is present, leading to a reduced and self-consistent set of jump conditions. 

The results presented here establish the Q-variable formulation as a
complete and consistent description of MHD discontinuities at the level
of conservation laws, while simultaneously exposing their underlying
characteristic structure. This provides a basis for further developments
in which the wave-aligned structure of the formalism can be exploited more
fully. In particular, a detailed classification of MHD discontinuities
within the Q-variable framework, including rotational discontinuities and
shock families, will be addressed in a separate study. In addition, the
application of the Q-variable formalism to Riemann problems and
characteristic-based numerical schemes will be explored in future work.

Although the present study is analytical, the formulation developed here
is well-suited for applications in both numerical modelling and the
interpretation of observations. By providing a wave-aligned
representation of the Rankine-Hugoniot conditions, the Q-variable
formalism offers a complementary perspective to the classical approach
and a natural framework for studying wave-shock interactions in
magnetised plasmas.

\section*{Funding}
T.V.H. acknowledges support from the KU Leuven grant STG/24/018.

T.V.D. was supported by a Senior Research Project (G088021N) of the FWO
Vlaanderen. Furthermore, TVD received financial support from the Flemish
Government under the long-term structural Methusalem funding program,
project SOUL: Stellar evolution in full glory, grant METH/24/012 at KU
Leuven. The research that led to these results was subsidised by the
Belgian Federal Science Policy Office through the contract
B2/223/P1/CLOSE-UP. It is also part of the DynaSun project and has thus
received funding under the Horizon Europe programme of the European
Union under grant agreement (no. 101131534). Views and opinions
expressed are however those of the author(s) only and do not necessarily
reflect those of the European Union and therefore the European Union
cannot be held responsible for them.
\section*{Declaration of interests}

The authors report no conflict of interest.

\bibliographystyle{apsrev4-1}
\bibliography{bibliography}
\end{document}